\documentclass[aps,twocolumn]{revtex4}
\usepackage{graphicx}
\usepackage{dcolumn}
\usepackage{bm}
\usepackage{amsmath}
\usepackage{amsfonts}
\usepackage{amssymb}
\begin{document}
\title{Classical  search algorithm with resonances in $\sqrt{N}$ cycles}
\author{A. Romanelli}
\altaffiliation[Corresponding author.]{{\it{E-mail:}}
alejo@fing.edu.uy}
\author{R. Donangelo}
\altaffiliation[Permanent address: ]{Instituto de F\'{\i}sica,\\
Universidade Federal do Rio de Janeiro\\
C.P. 68528, 21945-970 Rio de Janeiro,Brazil}

\affiliation{Instituto de F\'{\i}sica, Facultad de Ingenier\'{\i}a\\
Universidad de la Rep\'ublica\\ C.C. 30, C.P. 11000, Montevideo,
Uruguay}

\date{\today}

\begin{abstract}
 In this work we use the  wave equation to obtain  a classical
analog of the quantum search algorithm and we verify  that the
essence of  search algorithms resides in the establishment of
resonances between the initial and the serched  states. In
particular we show that, within a set of $N$ vibration modes,  it
is possible to excite the searched mode in a number of steps
proportional  to $\sqrt N$.\\
{\it{Keywords:}} search algorithm; quantum optics; quantum information\\
\end{abstract}

\pacs{PACS: 03.67.Lx; 72.15.Rn}

\maketitle
\section{Introduction}
It has been shown that a quantum search algorithm is able to locate a marked
item from an unsorted list of $N$ elements in a number of steps proportional
to $\sqrt{N}$, instead of proportional to $N$ as is the case for the usual
algorithms employed in classical computation. The most well studied quantum
search algorithm is the one due to Grover \cite{Chuang}, where the search is
performed by alternatively shifting the phase of the searched for state, and
amplifying its modulus. A continuous time version of the original Grover
algorithm has been described by Farhi and Gutmann \cite{Farhi}. In a recent
work \cite{alejo1}, we have presented an alternative continuous time quantum
search algorithm, in which the search set is taken from the set of
eigenvectors of a particular hamiltonian. The search is performed through the
application of a perturbation which leads to the appearance of a resonance
between the initial and the searched state. That study also provided a new
insight on the connections between discrete and continuous time search
algorithms. Our search algorithm can be implemented using any Hamiltonian with
a discrete energy spectrum. However, we would like to emphasize that the
possibility of establishing a resonance between two states is an intrinsic
property of oscillatory motions in general and not an exclusive property of
quantum mechanics as described through the Schr\"{o}dinger equation, as in our algorithm.

A link between quantum computation and classical optical waves has been well
established by several
authors\cite{Cerf,Roldan,Puentes,Pittman,Bhatta,Jennifer}. It is possible to
simulate the behavior of some simple quantum computers using classical optical
waves. Although the necessary hardware grows exponentially with the number of
qubits that are simulated, and thus these simulations are not efficient,
optical simulations could still be very useful. In fact, as some simulations
of quantum algorithms employ optical simulations, a classical analogue of such
a quantum search algorithm might be a valuable tool to test the functioning of
the optical system as a computer.

In this work we present a continuous time search algorithm, which is
controlled by a classical wave equation, showing explicitly, once again, that
the search algorithm is essentially a resonance phenomenon between the initial
and the searched states \cite{alejo1,Grover2}. The paper is organized as
follows: in the next section we briefly describe our quantum search algorithm.
Then, in section 3, we develop the search model using the ordinary wave
equation. In section 4 we present results obtained with this model. We end by
discussing these results and extracting conclusions, in the last section of
this work.

\section{Quantum search algorithm}

\label{quantum algorithm}We consider a continuous time quantum search
algorithm which is controlled by a time dependent Hamiltonian $H$. The
wavefunction $|\Psi(t)\rangle$ satisfies the Schr\"{o}dinger equation%

\begin{equation}
i\frac{\partial|\Psi(t)\rangle}{\partial t}=H\text{ }|\Psi(t)\rangle
,\label{Schrodinger}%
\end{equation}
where $H=H_{0}+V(t)$ and we have taken the Planck constant $\hbar=1$. Here
$H_{0}$ is a known nondegenerate time-independent Hamiltonian with a discrete
energy spectrum $\left\{  \varepsilon_{n}\right\}  $ and eigenstates $\left\{
|n\rangle\right\}  $. $V(t)$ is a time-dependent potential that shall be
defined below. We then consider a subset \textbf{N} of $\left\{
|n\rangle\right\}  ,$ formed by $N$ states, which will be our search set. Let
us call $|s\rangle$ the unknown searched state in \textbf{N} whose energy
$\varepsilon_{s}$ is given. We assume that it is the only state in \textbf{N}
with that value of the energy. Therefore, knowing $\varepsilon_{s}$ is, in our
algorithm, equivalent to the action of ``marking'' the searched state, in
Grover's algorithm. The potential $V$ that produces the coupling between the
initial and the searched states, is defined as \cite{alejo1}
\begin{equation}
V(t)=\left|  p\right\rangle \left\langle j\right|  \exp\left(  i\omega
_{sj}t\right)  +\left|  j\right\rangle \left\langle p\right|  \exp\left(
-i\omega_{sj}t\right)  \label{potencial0}%
\end{equation}
where the eigenstate $|j\rangle$, with eigenvalue $\varepsilon_{j}$, is the
initial state of the system and it is chosen so that it does not belong to the
subset \textbf{N}. Above, $\left|  p\right\rangle \equiv\frac{1}{\sqrt{N}%
}{\displaystyle
\sum\limits_{n\in{\mathbf{N}}}}|n\rangle$ is an unitary vector which can be
interpreted as the average of the set of vectors in \textbf{N}, and
$\omega_{sj}\equiv\varepsilon_{j}-\varepsilon_{s}$. This definition ensures
that the interaction potential is hermitian, that the transition probabilities
$W_{nj}\equiv\left|  \left\langle n|V(t)|j\right\rangle \right|  ^{2}=\frac
{1}{N}$, from state $|j\rangle$ to any state of the set \textbf{N} are all
equal, and finally, that the sum of the transition probabilities verifies
${\displaystyle\sum\limits_{n\in\mathbf{N}}}W_{nj}=1$. \begin{figure}[ptbh]
\begin{center}
\includegraphics[scale=0.38]{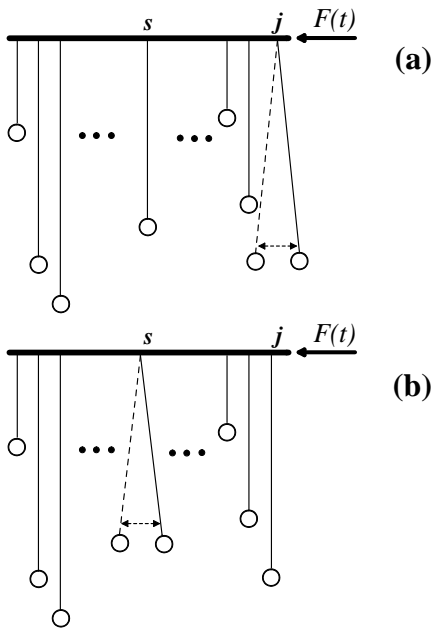}
\end{center}
\caption{(a) Initial configuration of the system: all $N$ pendulums at rest,
except pendulum $j$. (b) Situation after $\sqrt{N}$ cycles: most of the energy
was tranferred to the searched pendulum $s$.}%
\label{cero}%
\end{figure}The objective of the algorithm is to find the eigenvector
$|s\rangle$ whose transition energy from the initial state $|j\rangle$ is the
Bohr frequency $\omega_{sj}$. At this point we consider important to make a
digression to visualize the way the perturbation potential $V(t)$ operates on
the system. To better understand the process, let us consider a classical
system composed of $N$ pendulums of different lengths, and, consequently,
different oscillating frequencies. The search problem, illustrated in
fig.~\ref{cero}, consists of identifying the pendulum $s$ with a given
frequency of oscillation $\omega_{s}$. To perform the search, one suspends all
the pendulums from the same beam, and puts to oscillate one of the pendulums,
$j$. By applying a periodical force with frequency $\omega_{j}-\omega_{s}$,
which may be derived from a potential $V(t)$, the sought after pendulum $s$
receives, after some time, most of the oscillation energy of the pendulum $j$,
so it may be readily identified. This system, closely related to the one
proposed by Grover \cite{Grover2}, also has the property that the
identification is performed through the gradual exchange of energy between the
oscillators, without a net external energy input.

Returning to the quantum search problem, the wavefunction $|\Psi(t)\rangle$
appearing in Schr\"odinger's eq.~(\ref{Schrodinger}) may be expressed as an
expansion on the eigenstates $\{|n\rangle\}$ of $H_{0}$,
\begin{equation}
|\Psi(t)\rangle=\sum_{m}a_{m}(t)\exp\left(  -i\varepsilon_{m}t\right)
|m\rangle. \label{expansion}%
\end{equation}
The time dependent coefficients $a_{m}(t)$ have initial conditions
$a_{j}(0)=1$, $a_{m}(0)=0$ for all $m\neq j$. After solving the Schr\"{o}%
dinger equation, the probability distribution results in,
\begin{align}
P_{j}  &  \simeq\cos^{2}(\Omega\ t)\ ,\nonumber\\
P_{s}  &  \simeq\sin^{2}(\Omega\ t)\ ,\label{psearched}\\
P_{n}  &  \simeq0,\text{ }n\neq j\text{ and }n\neq s,\nonumber
\end{align}
where $\Omega=1/\sqrt{N}$. From these equations it is clear that for $\tau
=\pi/(2\Omega)$ a measurement yields the searched state with a probability
very close to one. This approach is valid as long as all the Bohr frequencies
satisfy $\omega_{nm}\gg\Omega$ and, in this case, our method behaves
qualitatively like Grover's.

\section{Search algorithm using classical waves}

\label{classical algorithm}In this section we build a classical search
algorithm which is controlled by a classical wave equation with an added
perturbation. The task of this algorithm is, starting from an initial
oscillation mode, to excite the searched mode, whose frequency is known, in
$O(\sqrt{N})$ steps. The unperturbed wavefunction $\Psi(x,t)$ satisfies the
ordinary wave equation in one spacial dimension,%
\begin{equation}
\frac{\partial^{2}\Psi(x,t)}{\partial t^{2}}=c^{2}\frac{\partial^{2}\Psi
(x,t)}{\partial x^{2}},\label{ondas1}%
\end{equation}
where $c$ is the constant wave velocity. This equation plays the same role as
the Schr\"{o}dinger equation in the algorithm shown in the previous section.
From eq. (\ref{ondas1}) we define the basis on which the algorithm is built.
Let us call $L$ the length of the region where the wave equation is solved.
Then the canonical basis is given by the modes $\exp(-i\omega t)\sin
(Kx)$\ where $K$ is the wave number and $\omega$ is the circular frequency,
connected through the dispersion relation $c=\omega/K$. In this case, the wave
number and the frequency are of the form $K_{n}=K_{0}n$ and $\omega_{n}$
$=\omega_{0}n$ respectively, where $n$ is an integer, $K_{0}=2\pi/L$ and
$\omega_{0}$ $=cK_{0}$.

As the analogue of the Schr\"{o}dinger equation eq.(\ref{Schrodinger}) with
the complete Hamiltonian $H=H_{0}+V(t)$ we consider the wave equation with a
perturbative force term,%

\begin{equation}
\frac{\partial^{2}\Psi(x,t)}{\partial t^{2}}=c^{2}\frac{\partial^{2}\Psi
(x,t)}{\partial x^{2}}+c\omega_{0}\frac{\partial\left(  V(t)T\Psi(x,t)\right)
}{\partial x}\label{ondas2}%
\end{equation}
where $V(t)$ is a external time depend perturbative potential. The solution of
eq. (\ref{ondas2}) can be expressed as a linear combination of the oscillation
modes, of eq. (\ref{ondas1})%
\begin{equation}
\Psi(x,t)=\sum_{n}a_{n}(t)\exp(-i\omega_{n}t)\sin(K_{n}x),\label{función1}%
\end{equation}
where the coefficients are%
\begin{equation}
a_{n}(t)=\exp(i\omega_{n}t)\frac{2}{L}\int\limits_{0}^{L}\Psi(x,t)\sin
(K_{n}x)\text{ }dx.\label{función2}%
\end{equation}
\ In order for the algorithm to work, the operator $T$ in eq.(\ref{ondas2}) is
defined as
\begin{equation}
T\Psi(x,t)=\sum_{n}a_{n}(t)\exp(-i\omega_{n})\cos(K_{n}x).\label{función3}%
\end{equation}

In order to obtain a greater connection with the previous section, we shall
now adopt the Dirac notation for the coordinate space $|x\rangle$, and for
momentum space $|n\rangle$. In this framework, the solution of the wave
function can be expressed as%
\begin{equation}
|\Psi\rangle=\sum_{n}a_{n}(t)\exp\left(  -i\omega_{n}t\right)  |n\rangle
,\label{coeficiente}%
\end{equation}
where we can write
\begin{equation}
a_{n}(t)=\frac{2\exp\left(  i\omega_{n}t\right)  }{L}\int\limits_{0}%
^{L}dx\left\langle n|x\right\rangle \left\langle x|\Psi\right\rangle
,\label{aes}%
\end{equation}
with $\left\langle n|x\right\rangle =\sin(K_{n}x)$.

Once again, the task of the algorithm is to excite the oscillation mode
$|s\rangle$ starting from the initial mode $|j\rangle$, with the transition
frequency $\omega_{sj}\equiv\omega_{j}-\omega_{s}$. To perform this task, we
choose as perturbation potential to excite the resonance between the initial
and searched states the classical analogue of the one given by
eq.(\ref{potencial0}). In our mixed notation, this time depended perturbation
is%
\begin{equation}
V(t)\Psi(x,t)\equiv\left\langle x|V(t)|\Psi\right\rangle =\sum_{n}\left\langle
n|V(t)|\Psi\right\rangle \sin\left(  K_{n}x\right)  ,\label{funcpoten}%
\end{equation}
where%
\begin{align}
\left\langle n|V(t)|\Psi\right\rangle  &  =\frac{1}{\sqrt{N}}\{\left(
1-\delta_{nj}\right)  a_{j}(t)\exp\left(  i\omega_{sj}t\right)  \label{braket}%
\\[0.01in]
&  +\delta_{nj}\exp\left(  -i\omega_{sj}t\right)  \sum_{l\in\mathbf{N}}%
a_{l}(t)\exp\left(  -i\omega_{l}t\right)  \}.\nonumber
\end{align}
The last equation has been obtained from eqs.(\ref{potencial0},\ref{expansion}%
). Using eqs.(\ref{función1},\ref{funcpoten},\ref{braket}) in eq.(\ref{ondas2}%
) and projecting the result on the mode $n$ the following set of equations for
the amplitudes $a_{n}(t)$ are obtained%
\begin{align}
\frac{d^{2}a_{n}(t)}{dt^{2}}-2i\omega_{n}\frac{da_{n}(t)}{dt} &
=-\frac{\omega_{0}}{\sqrt{N}}\{\omega_{n}\left(  1-\delta_{nj}\right)
a_{j}(t)\exp\left(  i\omega_{sn}t\right)  \label{dinamics}\\
&  +\omega_{j}\delta_{nj}\sum_{l\in\mathbf{N}}a_{l}(t)\exp\left(  i\omega
_{ls}t\right)  \}.\nonumber
\end{align}
These equations will be solved in the next section but meanwhile we address
the question of their qualitative behavior. The eqs.(\ref{dinamics}) involve
two time scales, a fast scale associated with the frequencies $\omega_{n}$,
and a slow scale associated with the amplitudes $a_{n}(t)$. In a time interval
larger than a characteristic time of the fast scale, the most important terms
are those that have a very small phase, as the others average to zero. In this
approximation the previous set of equations becomes
\begin{align}
\frac{da_{j}(t)}{dt} &  \simeq-i\Omega a_{s}(t),\nonumber\\
\frac{da_{s}(t)}{dt} &  \simeq-i\Omega a_{j}(t),\label{dinamics3}\\
\frac{da_{n}(t)}{dt} &  \simeq0\text{ for all }n\neq s,j;\nonumber
\end{align}
where $\Omega=\omega_{0}/(2\sqrt{N})$. These equations represent two coupled
oscillators where the coupling is established between the initial and the
searched states. Solving them with initial conditions $a_{j}(0)=1$,
$a_{s}(0)=0$ we obtain
\begin{align}
a_{j}(t) &  \simeq\cos(\Omega t),\nonumber\\
a_{s}(t) &  \simeq i\sin(\Omega t).\label{psearched}%
\end{align}
\begin{figure}[h]
\begin{center}
\includegraphics[scale=0.38]{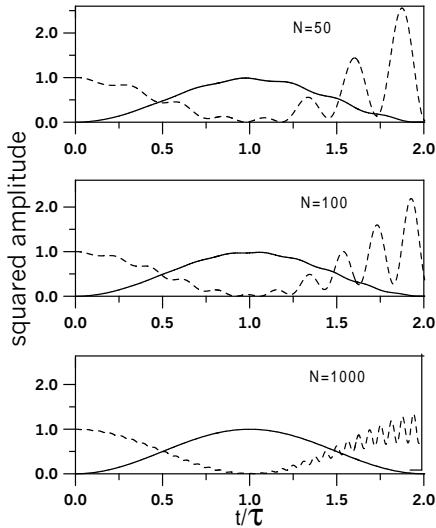}
\end{center}
\caption{Time evolution of the squared amplitude of the initial
state (dashed line) and the searched state (full line) for
different number of modes $N$.
The frequencies $\omega_{n}$ are taken with $n=1,2,3,...,N$.}%
\label{uno}%
\end{figure}
It is important to note that the previous approach is valid only if all the
frequencies verify $\omega_{n}\gg\Omega$. The amplitude of the
searched state is maximum at the time
$\tau\equiv\pi\sqrt{N}/\omega_{0}.$ This time is equivalent, in
the Grover algorithm, to the optimal time to perform the
measurement. Then, when the previous approximations are valid, our
method behaves qualitatively like Grover's. The time $\tau$ has a
different prefactor than the searching time of the Grover
algorithm,\cite{Chuang} but this difference is not important
because the resonance potential eq.(\ref{potencial0}) can be
adjusted through multiplication by a constant factor.

\section{Numerical results}
\begin{figure}[h]
\begin{center}
\includegraphics[scale=0.3]{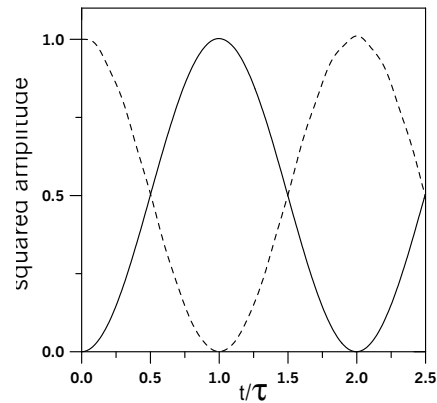}
\end{center}
\caption{ Squared amplitude of the initial state (dashed line) and
the searched state (full line) as a function of time for $N=100$
modes. The
frequencies $\omega_{n}$ are taken with $n=1,4,9,...,N^{2}$.\newline }%
\label{dos}%
\end{figure}
\label{sec:Numerical results} We have integrated numerically
eqs.(\ref{dinamics}) with initial conditions $a_{j}(0)=1$,
$a_{n}(0)=0$ for all $n\neq j$, and have verified that the
solutions so obtained are independent of the remaining initial
conditions for the derivatives $\frac{da_{j}(0)}{dt}$and
$\frac{da_{s}(0)}{dt}$. The calculations were performed using a
standard fourth order Runge-Kutta algorithm. Choosing an arbitrary
frequency for the searched mode, we follow the dynamics of the set
\textbf{N}. We have verified for several values of $N$ that the
most important coupling is between the initial and the searched
mode; other couplings being negligible as discussed in the
previous section. We worked with two types of sets \textbf{N, }in
the first set the mode frequencies $\omega_{n}=\omega _{0}n$ are
taken with $n=1,2,3,...,N$ and in the second set with
$n=1,4,9,...,N^{2}$. The square of the amplitude of the searched
and the initially loaded modes are shown for the first set in
Fig.\ref{uno}. The temporal evolution was normalized to the
characteristic time $\tau$. Each panel shows the square of the
amplitude for $N=50,100,1000$ modes respectively. From this
figure, it is clear that the amplitudes of the initial mode and
the searched mode alternate in time, as in the Grover algorithm.

The time at which the squared amplitude of the searched state is maximum and
its value is very near one agrees with our theoretical prediction $\tau$. The
agreement is much improved when $N$ increases because the condition
$\Omega<<\omega_{nm}$ is better satisfied as $\Omega\sim1/\sqrt{N}$. When
$t\sim2\tau$ the amplitude of the initial mode shows strong oscillations which
decrease as the inverse of the search set size. We also observe that the
frequency of these oscillations increases \ with $N$.

Fig.\ref{dos} shows the squared amplitude of the initial and the searched
modes as a function of time for the second set. In this case the calculation
agrees with the theoretical prediction for all $N>10$, and thus we have
displayed only the calculation for $N=100$.

The search algorithm dynamics for the two sets, are rather similar. The
differences observed are due to the properties of their frequency spectra.
While the frequency spectrum of the first set increases linearly with the mode
number, the frequency of the second set has a quadratic growth. Then in the
first set, an additional interference effect must be expected because many
more frequencies $\omega_{sn}$ have the same value for different values of $s$
and $n$ in the case of the first set than in the case of the second set. In
this last case the frequencies $\omega_{sn}$ behave pseudorandomly and the
additional interferences do not appear.

\section{Conclusions}

\label{sec:conclusion} We have developed a search algorithm using a classical
wave equation with perturbation. This algorithm behaves like the Grover
algorithm, in particular the optimal search time is proportional to the square
root of the size of the search set and the probability to find the searched
state oscillates periodically in time. The efficiency of this algorithm
depends on the spectral density of frequencies of the set where the search is
made. A larger separation between frequency levels maximizes the probability
of the searched state and allows for a better precision. Nevertheless the
classical implementation of this algorithm requires $N$ oscillation modes,
whereas in a quantum system the implementation will only need $\log_{2}$ $N$
qubits\cite{Grover2}.

Presently most of the attention is devoted to the development of quantum
computational devices. In particular, there have been several recent attempts
to employ optical simulations to run quantum algorithms
(\cite{Cerf,Bhatta,Puentes}). These studies suggest that such optical
implementations are possible for small systems with the presently available
technology. Here, we have developed a general classical analogue of a quantum
search algorithm which operates in the same way as the one implemented in the
quantum case through optical devices. Thus, our work not only helps to
understand the essence of the quantum algorithm, but also may serve as a
testing ground for practical implementation of optical computational systems.
Certainly much further experimental and theoretical work will be needed to
develop the practical applications of this algorithm.

We acknowledge useful comments made by S. Barreiro, A. Lezama and V.
Micenmacher, and financial support from PEDECIBA and PDT S/C/IF/54/5. R.D.
acknowledges partial financial support from the Brazilian National Research
Council (CNPq) and FAPERJ (Brazil). A.R and R.D. acknowledge financial support
from the \textit{Brazilian Millennium Institute for Quantum Information}.

\end{document}